\documentclass[12pt,preprint,useAMS,usenatbib,epsfig]{aastex}
\begin{document}

\title{Ice emission and the redshifts of submillimeter sources}

\date{}
\author{C. C. Dudley}
\affil{Formerly: Naval Research Laboratory, Remote Sensing Division, 4555 Overlook Ave. SW, Washington, DC 20375-5351,  USA}
\email{C.Christopher.Dudley.85@alum.dartmouth.org}
\author{M. Imanishi\altaffilmark{1}}
\affil{National Astronomical Observatory, Mitaka, Tokyo 181-8588, Japan}

\and

\author{P. R. Maloney}
\affil{Center for Astrophysics and Space Astronomy, University of Colorado, Bolder, CO 80309-0839, USA}

\altaffiltext{1}{Department of Astronomy, School of Science, Graduate University for Advanced Studies, Mitaka, Tokyo 181-8588}
\begin{abstract}
Observations at submillimeter wavelengths have revealed a population
of sources thought to be at relatively large redshifts.  The position
of the 850 $\mu$m passband on the Rayleigh-Jeans portion of the Planck
function leads to a maximum redshift estimate of $z\sim$4.5 since
sources will not retain their redshift independent brightness close to
the peak of the Planck function and thus drop out of surveys.  Here we
review evidence that ice absorption is present in the spectra of local
ultraluminous infrared galaxies which are often taken as analogs for
the 850 $\mu$m source population.  We consider the implication of this
absorption for ice induced spectral structure at far infrared
wavelengths and present marginal astronomical evidence that amorphous
ice may have a feature similar to crystalline ice near 150 $\mu$m.
Recent corroborative laboratory evidence is supportive of this
conclusion.  It is argued that early metal enrichment by pair
instability SN may lead to a high ice content relative to refractory
dust at high redshift and a fairly robust detection of ice emission in
a $z=6.42$ quasar is presented. It is further shown that ice emission
is needed to understand the 450 $\mu$m sources observed in the GOODS-N
field. We are thus encouraged to apply far infrared ice emission
models to the available observations of HDF 850.1, the brightest
submillimeter source in the {\it Hubble Deep Field}.  We suggest that
a redshift as large as 13 may need to be considered for this source,
nearly a factor of three above the usual top estimate.  Inclusion of
the possibility of far infrared ice emission in the spectral energy
distributions of model sources generally broadens the range of
redshifts to be considered for submillimeter sources compared to
models without ice emission.  
\end{abstract}


\section{Introduction}

Sources discovered in surveys undertaken using 450/850 $\mu$m SCUBA
camera mounted on the JCMT have a high number density per unit solid
angle ($\sim$ 10$^4$ deg$^{-2}$; S$_{850} >$ 1 mJy; \citealt{dave};
\citealt{wang}) and are thought to trace the dust emission from
sources at a range of redshifts.  Sources brighter than 1 mJy which
have Arp 220-like far infrared (FIR) spectral energy distributions
(SEDs) are ultraluminous in the $h_{100}=0.71$, $\Omega_m=0.27$, and
$\Omega_\Lambda=0.73$ cosmology consistent with cosmic microwave
background (CMB) and Type 1a SNe observations (\citealt{sperg};
\citealt{nolta}; \citealt{riess}) unless they are at redshifts $<$ 1
so that, thus far, submillimeter surveys do not detect ``normal''
galaxies at high redshift.  This should be the case even for
gravitational lens amplification up to a factor of 4 for 2 mJy limit
surveys.

It is therefore a sensible approach to consider the properties of
local ultraluminous infrared galaxies (ULIRGs; \citealt{daveandfelix})
as a guide to understanding sources detected in submillimeter surveys
with the proviso that the high redshift universe may not conserve
these properties.  It has already been noted that the number density of ULIRGs
increases with redshift at a rate much larger than can be explained
using geometric considerations alone (\citealt{kim}) and deep field
observations using the {\it Hubble Space Telescope} have found quite obvious
morphological evolution at higher redshifts compared to local
conditions ({\it e.g.}  \citealt{brinchmann}).  While these particular
evidences of evolution as a function of redshift do not directly speak
to ULIRG SED evolution, they do suggest that this proviso cannot be
lightly discounted.

Ice is present in the dense ISM of the Galaxy.  It is
observed to occur above a threshold depth of A$_V\sim$ 3
within molecular clouds but is absent on the cloud edges and
in the diffuse (atomic/ionized) ISM (\citealt{whittet}).
Because ice is not exposed to star light, far infrared (FIR) emission
from ice is not expected from quiescent clouds.  The FIR
emission from these ice bearing clouds arises from the first
few ice-free A$_V$.  To observe ice emission in the Galaxy,
centrally heated sources are needed.  If mid-infrared (MIR) radiation can
illuminate the ice bearing grains so that they may be
heated, then the FIR emission will be dominated by ice if
the ice mantles provide a moderate fraction of the total
grain volume.  This FIR dominance is owing to greater FIR
opacity of ice relative to silicate or carbonaceous grains.
In the models presented by \cite{aannestad}, a mantle-to-core
volume ratio of 0.33 doubles the 60 $\mu$m opacity: a 10\%
increase in grain radius doubles the opacity!

Protostars in the Galaxy which are centrally heated turn out
to be good sources for observing FIR ice emission ({\it
e.g.} \citealt{dartois}).  And, there is mounting evidence
that a significant fraction of local ULIRGs are also
centrally heated (\citealt{dww}; \citealt{soifer};
\citealt{iandd}; \citealt{masa}; \citealt{idm}; \citealt{spoon06})
so that the FIR emission is a result of MIR heating.  Thus,
if ice is present in the FIR emitting regions, its effects on
the FIR opacity may be important.  This situation differs
from what occurs in lower luminosity starburst galaxies
where the FIR emission arises from the surfaces of molecular
clouds which are heated by optical and far UV rather than MIR
light.  The interiors of the clouds, where ice presumably
exists, are not strongly heated by the mainly FIR emission 
arising from their surfaces. 

If submillimeter galaxies are similar to ULIRGs in more than 
just their luminosity, then they may also sometimes have FIR
emission that is powered by MIR light.  For redshift
estimation the question of the similarity between
submillimeter galaxies and local ULIRGs is as important as the
differences that distinguish ULIRGs from their lower
luminosity starburst counterparts.   

Attempts to estimate redshifts for the submillimeter sources
based on their radio fluxes could be guided by local ULIRG
samples.  For the 13 ULIRGs brighter than 5.24 Jy at 60
$\mu$m (\citealt{bgs}; \citealt{ebgs}) available to the VLA,
the mean logarithm of the ratio of FIR-to-20 cm radio flux
is 0.22 dex larger than $q$=2.43$\pm$0.19 (\citealt{chyt})
for starburst galaxies when calculated using radio
measurements reported by \cite{chyt} and \cite{min}.  While
the mean differs at only the 1$\sigma$ level, the dispersion
of $q$ is a factor 1.6 (in dex) larger than for the lower
luminosity star forming galaxies.  Using a different
luminosity binning, \cite{george} report a similar trend in
increased dispersion with increasing luminosity.  The
implication for redshift estimation is that the range of
likely redshifts would increase when using ULIRG properties
rather than starburst properties. And, a trend away from
what is seen in local ULIRGs, say through more powerful
starbursts at earlier times, could have complex implications
for estimating redshifts of submillimeter galaxies using
radio observations.

However, broad limits on the redshift distribution derive
from submillimeter observations alone where the relative
infrequency of detecting 450 $\mu$m emission leads to (weak)
lower limits on the redshifts while the detections
themselves at 850 $\mu$m lead to upper limits since, baring
extraordinary luminosities, the turnover from the
Rayleigh-Jeans-like portion of the SED at high redshift
leads to source fluxes dropping below survey detection
limits.

Given the large potential effect of ice on the FIR opacity
of submillimeter galaxies, we first review the evidence for
the presence of ice in centrally heated ULIRGs and estimate
an expected FIR SED for a case where FIR ice opacity can be
expected to be dominant based on a rough abundance analysis.
We then search for FIR ice spectral features in the
available data on ULIRGs.  We discuss why a high relative
abundance of ice during the epoch of reionization might be
expected and present somewhat firmer evidence for FIR ice
emission in the $z=6.42$ source QSO J1148+5251 than can yet
be demonstrated for ULIRGs.  We then take the submillimeter
galaxy HDF 850.1 as an example to show how FIR ice emission
allows a very large redshift estimate.  

\section{Ice in ULIRGs}

\subsection{Absorption}

Observations of strong absorption features from refractory
dust in ULIRGs trace dust that is cooler than the
temperature of a blackbody emitting at the wavelength of the
feature: $\sim$ 300 K for the 10 $\mu$m silicate feature and
$\sim$ 150 K for the 20 $\mu$m silicate feature for example.
Thus, 20 $\mu$m silicate features are weak in ULIRGs
compared to 10 $\mu$m features owing to the dust radial
temperature gradient. Ice absorption, on the other hand,
should only occur in the cooler outer regions of the dust
envelope and so may show the same relative feature strengths
found in the laboratory unless the absorbing covering factor
is a strong function of the emitting source's wavelength
dependent size (buried source radius $r\approx0.8({L\over
10^{12} L_\odot})^{0.5}({\lambda \over 3 \mu m})^2$ pc where
T has been eliminated between the Wein Displacement and
Stefan-Boltzmann Laws, $L$ is the portion of IR luminosity
corrected for extended star formation, $\lambda$ runs from 2
$\mu$m to the lesser of either the wavelength where the
source is no longer optically thick or the peak FIR
wavelength and scattering is neglected).  MIR ice absorption
is therefore consistent with a portion of the FIR emission
being powered by the MIR rather than the optical and UV
portion of the electromagnetic spectrum.  However, it is
necessary to be certain that the reported absorption is
indeed owing to ice.  As an example, \cite{spoon} claim ice
absorption in 18 galaxies based on 6 $\mu$m absorption
including IRAS 17208-0014 the third brightest ULIRG.
However OH bond bending needn't occur in ice alone.  The
absence of 3 $\mu$m absorption in the spectra of the dust
continuum dominated sources Mkn 231 and IRAS 05189-2524
presented by \cite{spoon}, where veiling by starlight is
unimportant, suggests that an identification of water ice
may be premature for these sources and refractory materials
such as those studied by \cite{greenberg} may provide a more
convincing model.  An apparent 6 $\mu$m dip in the spectrum
of Mkn 231 plotted by \cite{weedman} similar in amplitude to
the 6.2 $\mu$m hump for which they report a measurement
tends to confirm the \cite{spoon} claim of 6 $\mu$m
absorption in this source. The spectrum of Arp 220 may be
consistent with a contribution of 12.5 $\mu$m libration mode
absorption, supportive of the \cite{spoon} identification of
ice absorption in this source but, until recently
(\citealt{idm}; \citealt{guido}; \citealt{masa06}), the
only well confirmed case of ice absorption among ULIRGs in
their sample was that of IRAS 00188-0856 where ice
absorption at 3 $\mu$m confirms the identification
(\citealt{masa}).  See \cite{idmet} for further well
confirmed ice identifications.

To illustrate this we present Fig. 1 which employs a
slightly detailed model to explain the absorption features
observed between 2.7 and 12.4 $\mu$m.  The upper panel
reproduces the 3 $\mu$m spectrum reported by \cite{masa}
together with an absorption model composed of amorphous ice
and refractory carbonaceous absorption applied to a
power-law continuum, and veiled by some starburst-like
emission.  The model is then applied in the lower panel with
some adjustment made for expected reduced refractory
absorption at longer wavelengths owing to a presumed dust
temperature gradient.  It has not been possible to fit the
entire 3 $\mu$m absorption without using ice.  Thus, the
absorption at 6 $\mu$m also is partly attributed to water
ice.  With these data, 12.5 $\mu$m water ice absorption,
while included in the model, does not seem to be required so
that one may conclude that two features, and particularly
the degeneracy breaking short wavelength edge of the 3
$\mu$m absorption contribute to the fairly robust
identification of ice in IRAS 00188-0856.  We note that the 
need for strong silicate absorption in this source is confirmed 
by the clear presence of 18 $\mu$m silicate absorption (\citealt{idmet})

Thus, ice absorption could play a significant role in ULIRGs, but the
evidence is not yet as strong as that showing that the MIR emission is
often absorbed by refractory dust.  Evidence for or against FIR ice
emission or absorption in ULIRGs is in worse shape still.  This is
primarily owing to a lack of observational data.  In Fig. 2 we present
archival ISO LWS spectra of Mkn 231 (filled diamonds) and IRAS
17208-0014 (filled circles) which show spectral structure in the
shortest wavelength detector band SW 1. This structure is compared to
the spectrum of the icy source HD 161796 (\citealt{hoog}) transformed
to absorption.  For Mkn 231, as just noted, the absence of 3 $\mu$m
ice absorption would lead to somewhat strained geometric arguments to
allow acceptance of the \cite{spoon} identification of the 6 $\mu$m
absorption, but once made, such arguments would likely allow
absorption at even longer wavelengths.  A complex geometry has already
been invoked for this source on a smaller physical scale to explain
its low X-ray flux (\citealt{phil}).  For IRAS 17208-0014, the present
spectrum would tend to confirm the \cite{spoon} identification of the
6 $\mu$m absorption as do 3 $\mu$m spectra (\citealt{guido};
\citealt{masa06}).  For either source, the presence of crystalline ice
implied by the possible 43 $\mu$m absorption would imply a low cosmic
ray abundance at least in the region where the absorption is produced
since ion bombardment erases crystalline structure in cold ices
(\citealt{moore}).  The inset in Fig. 2 shows the relationship between
the wavelength of the 43 $\mu$m peak and ice temperature measured by
\cite{smith} which we have invoked to shift the ice spectrum derived
from HD 161796, an evolved ice forming star (\citealt{hoog}), beyond
the corrections made for the source redshifts to better match the
spectra.  Inversion of these corrections would imply temperatures at
which amorphous ice may exist and would be preferentially
formed. Constraints on cosmic ray abundance would need account for the
rate at which crystalline ice is introduced into the absorbing region
either through (substantial) changes in luminosity of the source which
may allow the annealing of amorphous ice to crystalline ice, processing
of ice by local heat sources in the region, or the migration of
crystalline ice into the region.  The discovery of substantial radial
molecular gas motion in the ULIRG IRAS 08572+3915 (\citealt{tom})
suggests the third possibility needs close scrutiny.  Since the SW1
detector of LWS is problematic, these sources must be independently
re-observed and the potential features confirmed before fully
accepting that ice is present in these sources.

\subsection{Millimeter morphology}

We have examined the LWS SW1 spectra of a two other ULIRGS, Arp 220
and Mkn 273, finding no feature consistent with crystalline ice
absorption.  In the case of Arp 220, the FIR spectrum appears to be
optically thick (\citealt{gonzalez}).  There is every reason to think
ice should be present owing to the source temperature, the clear
presence of water vapor, as well as supportive MIR observations
already mentioned.  Here, ice may help to resolve an issue raised by
\cite{soifer99} with regard to FIR sources that are optically thick
showing any detectable emission in the MIR owing to overwhelmingly
large implied MIR opacity.  Since ice strongly enhances the FIR
opacity compared to refractory dust, a given FIR optical depth can be
achieved with less dust, up to a factor of seven in the models
computed by \cite{aannestad}.  This effect can make the issue raised
by \cite{soifer99} less troubling, though other approaches such as
examining the effects of the dust density distribution ({\it e.g.}
\citealt{nenkova}) may also need to be explored as well.  Ice may also
have the effect of making millimeter continuum size estimates smaller
than what might be expected for refractory dust that is optically
thick in the FIR since the opacity of ice mantled grains is lower than
that of bare grains at millimeter wavelengths.  This could yield a
cuspy morphology in the millimeter where one would usually expect a
relatively more flattened morphology owing to the cooler outer regions
being relatively brighter in the millimeter under the assumption of
higher refractory dust opacity.

\subsection{FIR ice spectrum}

The infrared laboratory spectrum of crystalline water ice
has been well known for some time (\citealt{bertie}) and
differences between the spectrum of crystalline and
amorphous ice have been exploited to aid understanding of
astronomical sources ({\it e.g.}  \citealt{dartois}).  However
owing to the low thermal conductivity of amorphous ice,
preparation of laboratory samples, via vapor deposition,
that are thick enough to conduct investigations at
wavelengths longer than 100 $\mu$m has been too difficult
until recently.  In particular it has been unknown whether
or not amorphous ice has a feature corresponding to the
broad feature centered near 150 $\mu$m identified by
\cite{bertie} as a phonon mode in crystalline ice.  Since
nature provides sources rich in amorphous ice, it may be
that such a comparison could be made through observations of
these sources.  The spectral coverage of the ISO LWS
spectrometer extended to $\sim$ 200 $\mu$m, too short to
fully cover the broad 150 $\mu$m crystalline ice feature.
Comparison between the ISO LWS data presented by
\cite{dartois} for RAFGL 7009S, a deeply embedded
ultracompact \ion{H}{2} region, and the submillimeter
observations provided by \cite{mccutcheon} yield a modified
optically thin blackbody parameter set of T=27 K,
$\beta$=2.1 (where the dust emissivity is proportional to
$\lambda^{-\beta}$) when attempting to the match data near
190 $\mu$m.  The large value of $\beta$ is suggestive of the
presence of a feature similar to that of crystalline ice in
this amorphous ice dominated source.  However, full
spectroscopic confirmation is needed to decide if the long
wavelength departure of the data from the model of
\cite{dartois} indicates the existence of an amorphous ice
feature analogous to the 150 $\mu$m crystalline ice feature.

Laboratory work by \cite{curtis} has explored the optical
properties of amorphous ice between 15 and 200 $\mu$m.
There is little question that most of the ice in the thick
films used by \cite{curtis} deposited at $T\leq 126$ K
differs from ice deposited above this temperature since the
wing of the libration band is clearly shifted so it seems
safe to conclude that amorphous ice, clearly present in the
thinner samples, is also largely present in the thicker
samples despite issues related to the low thermal
conductivity of amorphous ice which hampers the (rapid)
growth of thick samples.  \cite{perry} have also had recent
success in growing thick samples of amorphous ice using low
deposition rates.  \cite{schober} take a different approach:
annealing samples of high density amorphous ice into the low
density amorphous ice found in space.  As with the LWS
spectroscopy, the new amorphous ice optical constants
(\citealt{curtis}) do not extend to the wavelength regime
(300 $\mu$m) where crystalline ice shows absorption whose
wavelength dependence is described by $\lambda^{-4}$
(\citealt{bertie}).  Unlike natural sources, however, where
temperature and grain size effects make full spectral
coverage of a broad feature nearly absolutely necessary to
ascertain its reality, the controlled laboratory conditions
allow a reasonable extrapolation to be considered reliable.
The finding of \cite{curtis} that the imaginary part of the
refractive indices of amorphous and crystalline ice are very
similar between 100 and 200 $\mu$m would appear to be
consistent with the suggestion of \cite{schober} that the
order in amorphous ice indicated by sharp features in
inelastic X-ray scattering data may be owing to a hydrogen
network that is also present in crystalline ice.  If this
order is indeed present, then a shared spectral structure
for the interaction of (FIR) photons with phonons between
the two types would allow a first order substitution of
crystalline ice optical constants for the unknown values for
amorphous ice beyond 200 $\mu$m.

\subsection{FIR ice emission}

Turning to photometry, and returning to ULIRGs, if we accept the
\cite{spoon} classification scheme then the spectrum of IRAS
14348-1447 as reported by \cite{genzel} or \cite{charmandaris} must be
considered Class II, that is 6 $\mu$m ``ice'' absorption is present
and partially filled in by 6.2 $\mu$m polycyclic aromatic hydrocarbon
emission.  This source occupies an extreme position in the dust
emissivity ($\beta$) vs. 100 $\mu$m optical depth ($\tau_{100}$)
diagram given by \cite{klaas} with both high $\tau_{100}$ (5) and
$\beta$ (2).  Since the effects of optical depth and emissivity are
intertwined, a lower optical depth can imply a larger value of
$\beta$.  For example, $\beta =2.4$ and $\tau_{100}=0.1$ does about as
well as the \cite{klaas} parameters for data longward of 100 $\mu$m.
However, values of $\beta$ larger than 2 are not generally found
except where ice is thought to be present ({\it e.g.}  \citealt{lis})
or dust temperatures are lower than found in ULIRGs (67 K for the
\cite{klaas} fit or 26 K at the lower optical depth and larger $\beta$
tested here).  In Fig. 3 we compare the fit proposed for IRAS
14348-1447 by \cite{klaas} with an alternative model.  The model shown
by the dashed line is that of \cite{klaas}, while that shown by the
thick solid line is a core-mantle grain model taken from
\cite{aannestad} but modified to have amorphous ice features in the
following way: First, opacity data from \cite{smith} for crystalline
ice cooled to 70 K was matched to the \cite{aannestad} curve for
V$_m$/V$_c$=0.95 by inducing a $\nu^{-1}$ dependence to approximate
the Mie calculation results along with the addition of an underlying
continuum opacity. Second, amorphous ice opacity data
(\citealt{smith}; deposited at 10 K and warmed to 70 K) was substituted
between 30 and 70 $\mu$m based on this matching after adjustment for
the differing sample thicknesses.  This opacity curve was used to
calculate a model with 16 and 71 K components.  If we take temperature
and optical depth as free parameters for the \cite{klaas} fit (2
parameters; normalization fixed by IRAS) and two temperatures, two
optical depths, and the relative contribution at 90 $\mu$m of these two
components as 5 free parameters for the present model then we may
estimate how well the two models represent the data if we know
something about the errors in the data.  The statistical errors for
the data are thought to be small but the absolute calibration error
could be large, $\sim$30\%.  What is important for comparing models is
the relative accuracy from filter to filter.  We have thus assumed
13\% errors for the ISOPHOT data to give a filter-to-filter accuracy
of 20\% (\citealt{klaas00}).  The fit of \cite{klaas} for data
longward of 20 $\mu$m is unacceptable (P($\chi^2$,$\nu$)=0.02) under
our assumption about the errors.  A marginally acceptable fit
(P($\chi^2$,$\nu$)=0.4) is found for the amorphous ice model.  Detailed
inspection of the ice fit suggests that the 150 $\mu$m ice feature is
not accounting for all of structure in the data.  The thin solid line
(V$_m$/V$_c$=0.33, T=28.8 K; $\tau_{90}=0.2$ and V$_m$/V$_c$=0.95, T=80.5 K,
$\tau_{90}=0.08$ crystalline ice; onion-skin model with warm, T=180 K, 
refractory dust in the interior) shows that this might be
possible.  However, our method of substituting amorphous for
crystalline ice is unlikely to work well for the thin mantle used
here.  Given the assumption regarding the data errors made here, it is
not possible to make a clear claim that ice emission is detected in
this source.  Assuming larger errors could make both low free
parameter models acceptable while reducing the errors could make the
simpler ice model unacceptable while the more complex model cannot be
evaluated via reduced $\chi^2$.  Clearly, {\it Spitzer} MIPS SED mode
observations could be definitive since the strong 70 $\mu$m curvature
in the ice models cannot be produced with standard refractory dust.

Thus, both astronomical and laboratory far infrared data
leave questions about ice unanswered.  Crystalline ice has a
feature near 150 $\mu$m which has yet to be completely
compared with amorphous ice.  Comparison between crystalline
and amorphous ice has been made near 40 and 60 $\mu$m and
they are quite distinct whereas they are similar between 100
and 200 $\mu$m. Beyond 200 $\mu$m questions remain which are
presently answered by extrapolation of the laboratory data
and hints from the astronomical data.  Substituting
crystalline ice for more common (in space) amorphous ice to
consider wavelengths longer than 100 $\mu$m could add to
(systematic) uncertainties in photometric redshift
estimation that can not be well addressed without further
observations or measurements.  Regardless of the remaining
issues concerning the 150 $\mu$m feature, for the main
purpose of this work it is prudent to consider the more
extreme consequences of crystalline ice emission on redshift
estimation which are examined after the next section.

\section{Ice at high redshift}

\subsection{Pair instability SNe enrichment}

Just as interplanetary grains may preserve a record of the
sources of ISM enrichment occurring at what is presently
observed to be $z=0.43$, the solid phase of the ISM at high
redshift may also record the conditions of the earliest
enrichment.  Particularly for sources that contribute to the
reionization of the universe ($7<z<14$; \citealt{spergel06}),
scenarios which attribute substantial enrichment to pair
instability SNe early on may
leave a mark on the solid phase of the ISM which could
linger through some subsequent processing.  A chief
attribute of pair instability SN enrichment is an early high
abundance of oxygen relative to other metals that usually 
deplete to the the solid phase of the ISM.

For example, a 186 M$_\odot$ progenitor yields number
abundance ratios of O:Si:C:S:Mg:$^{56}$Ni = 126:29:15:10:8:1
(\citealt{heger}) which, after formation of CO (29 O
removed), olivine (16 O and 4 Si removed), silica (50 O
removed), and SO$_2$ (20 O removed), leaves 9\% of the oxygen
available to form OH and H$_2$O.  For a 100 M$_\odot$
progenitor, O:C:Mg=48:9:1 with little Si, S or $^{56}$Ni
leaving 80\% of the oxygen available to combine with
hydrogen assuming the refractory solid state is MgO.  In the
absence of $^{56}$Ni, the formation of solids should be
enhanced since it may commence at a higher density in the
SNR.  These estimates for the available oxygen are
essentially lower limits, more complete treatments which
consider the possibility that the ejecta are unmixed
(\citealt{nozawa}; \citealt{schneider}; allowing the
formation of graphite for example) would leave an even
larger fraction of oxygen available.  Thus, we might expect
grains with large mantle-to-core volume ratios to result
from a period of pair instability SN enrichment.  The
continued presence of these ice grains would depend on the rate
of reprocessing of this initial solid phase in the ISM.

\subsection{Ice emission in a high redshift quasar}

The Gunn-Peterson trough in QSO J1148+5251 ($z=6.42$;
\citealt{berto2}) suggests that this source may be
participating in the last phases of reionization
(\citealt{fan03}).  That the quasar exists indicates that
the source has been contributing to reionization for some
time and we may take the beginning of reionization as an
estimate of it's age ($\sim$500 Myr) which gives a high
average accretion rate for a blackhole mass of $3\times
10^9$ M$_\odot$ (\citealt{willott}).  Thus, it seems
plausible that pair instability SN enriched material with a
high ice content is being supplied toward the center of this
activity as the blackhole and presumably its stellar system
grow in mass.  On the other hand, ice also forms in the more
evolved ISM of the Galaxy so an indication of the presence
of ice needn't be fully attributed to ice rich regions that
might be formed by pair instability SNe.

Fig. 4 shows an indication that the rest frame FIR emission from QSO
J1148+5251 is better explained by ice than refractory dust emission.
Three models are compared to photometric observations: 1) a refractory
model (dashed line T=55 K; $\tau_{60}=0.3$; $\beta=2$) is similar to
that given by \cite{beelen}, 2) a simple ice model (dot-dashed
line; T=43 K; $\tau_{60}=0.4$, V$_m$/V$_c$=5.6 amorphous), and 3) a
more complex ice model described below (solid line).

In Fig. 4, it is not possible to formally access the
goodness of fit through a reduced $\chi_\nu^2$ calculation
even for the refractory dust model (dashed line) since the
number of free parameters (normalization, temperature,
optical depth, and opacity index) is the same as the number
of points to be fit.  If one were to assume $\beta=2$ owing
to insufficient time since the first enrichment to produce
large grains then $\chi_\nu^2=10.7$ which strongly rules out
(P($\chi,\nu$)=0.001) this model.  That the fit is done by
eye probably does not affect this assessment but a failure to
account for systematic errors could since different groups
give 850 $\mu$m photometric estimates using the same data
that vary by as much as the refractory dust model differs
from the present data.  By selecting a particular
mantle-to-core volume ratio (5.6) we may comparably provide
a single degree of freedom and obtain $\chi_\nu^2=3.9$
(dot-dashed line) which would indicate that the model is not
fully ruled out by the data (P($\chi,\nu$)=0.05) and is an
improvement on the refractory dust model, again with the
proviso that we have only considered statistical errors.
More elaborate (multi-temperature) refractory dust or ice
mantled grain models can fit the detections essentially
perfectly and are useful to calculate as they can indicate
what further observations would be decisive.  The solid line
in Fig. 4 is such a model where T=71 K; $\tau_{60}=0.4$
ice-mantled dust is obscured by T=24 K; $\tau_{60}=0.4$
ice-mantled dust.  A similar model using $\beta=2$
refractory dust (not plotted; T=300 K; $\tau_{60}=0.1$ is
obscured by T=18 K; $\tau_{60}=1$ dust) exceeds the plotted
400 $\mu$m upper limit (arrow in Fig. 4; 0.39 mJy 3$\sigma$;
\citealt{berto1}) by a factor of 1.9 and is unphysical in a
number of ways including that the cool component is cooler
than the CMB at $z=6.42$. The ice models distinguish
themselves from the plotted refractory dust model by being
fainter by at least a factor of 4 at rest wavelength 400
$\mu$m (3 mm observed frame). The present upper limit does
not distinguish these cases.  Such a measurement should be
possible using CARMA.  Determining the composition of the
emitting material can also affect estimates of the dust mass
and thus estimates of the amount of past enrichment (and
star formation) in the QSO J1148+5251 system.  For the
present ice model, 7 times as much refractory dust would be
needed to produce the same FIR opacity so that a lower level
of total enrichment might be accommodated.

Finally, at high redshift, where theory suggests that we
might most anticipate it owing to expectations about early ISM
enrichment, we find evidence for FIR ice emission which,
while still debatable, seems moderately persuasive and which
can be relatively easily checked with further observations
using existing telescopes.  Now we may turn to our main
theme.  We have employed amorphous ice to consider IRAS
14348-1447 and QSO J1148+5251, but now we consider
crystalline ice both because its effect on redshift
estimation is the most dramatic and because ice may form at
fairly warm temperatures in the SNRs associated with pair
instability SNe.

\subsection{Ice emission in 450 $\mu$m submillimeter sources}

A set of sources that could be at high redshift and which demonstrate
the effects of ice are reported by \cite{pope}.  They are detected at
450 $\mu$m but not at 850 $\mu$m.  We consider the 7 sources tabulated
by \cite{pope} and two sources from the original list of
\cite{borys03} which were recovered above 3.5 $\sigma$ by \cite{pope}.
These have drawn our attention because the reported 850 $\mu$m upper
limits indicate large values of $\beta$ (\citealt{borys04}; based on
original list).  Large values of $\beta$ can indicate the presence of
ice emission.  If we generate 3$\sigma$ lower limits to the 450-to-850
$\mu$m flux density ratios using the 450 $\mu$m measurements less
3/$\sqrt(2)\times\sigma_{450}$ together with
3/$\sqrt(2)\times\sigma_{850}$ 850 $\mu$m upper limits then the
lowest lower limit is 8.7 and is only one of two that can be
consistent with $\beta<2$ dust.  This method of calculating limits can
be more conservative than calculating the ratio of the 450 measurement
to the 850 $\mu$m 3$\sigma$ upper limit (giving 13.4 in this case) but
its extension to 2$\sigma$ should be viewed with caution since
detections at the 2.9 $\sigma$ level are customarily reported as
3$\sigma$ limits.  In such situations, when data with comparable
signal-to-noise ratios become available, the ``2$\sigma$ extension''
may be violated much more frequently than the name implies if many of
these near misses lurk in the original data.  However,
\cite{pope} report no detection from applying the method of \cite{ch}
(stacking) to the 850 $\mu$m data at the positions of the 450 $\mu$m
sources they detect above 3.5$\sigma$.  In five cases where upper
limits are calculated below we substitute the ratio of the 450 $\mu$m
3$\sigma$ upper limit-to-the 850 measurement when it gives a more
conservative value.

In Fig. 5 we plot, as a function of redshift, the 450-to-850 $\mu$m
flux density ratio calculated for a series of SEDs.  The dashed line
is for a model with T=60 K, $\beta$=2, and $\tau_{90}$=2.  On this line
are plotted at $z=0.9$ and $z=0.5$ the 3$\sigma$ lower limits (filled
circle and diamond) for SMM J123603+620942 and J123631+620657.  The
arrows in the figure extend to the 2$\sigma$ lower limits.  In these
cases the argument given by \cite{borys03} for $z<1$ may be applied
directly, though tenuously, at the 3$\sigma$ level with no assumptions
about ice but this is not the case for the 7 remaining lower limits.
There may be some difficulty however accommodating these two sources at
these low redshifts unless sources with 0.2--2 Jy level flux densities
are present at 70 $\mu$m while \cite{frayer} find about four
$\sim30$ mJy sources at 70 $\mu$m in this region and we are unaware of
any IRAS sources in this field.  At these low redshifts, it is a
losing proposition to attempt to increase the dust temperature to
increase the redshift to avoid the 70 $\mu$m constraint in an effort
to avoid requiring ice emission.  When ice is invoked, then these and
the remaining lower limits may all be accommodated but the range of
consistent redshifts expands.  The remaining curves in Fig. 5
correspond to increasing ice temperature.  The dot-dashed curve uses
an SED similar to that shown in Fig. 1 (T=30 K), the triple-dot-dashed
curve has T=50 K as in Fig. 7., and the solid curve has T=150 K;
$\tau_{90}=0.04$ and V$_m$/V$_c$=5.6 for all ice models.  For the
solid curve, ambiguity between amorphous and crystalline ice is
physically resolved in favor of crystalline ice.  The SMM
J123603+620942 and J123631+620657 limits are replotted on the solid
curve giving a redshift limit $z<10.2$.  This limit is little changed
if the maximum possible ice temperature (170 K) is used.  All the
3$\sigma$ limits can be accommodated at $z=9.5$ but only SMM
J123603+620942 and J123631+620657 can also be accommodated there at the
2$\sigma$ level.  SMM J123657+622033 (filled square) shows this
clearly where it is plotted at $z=9.6$.  The 2$\sigma$ level for SMM
J133747+621600 (filled clover leaf) also constrains the ice
temperature to be T$\gtrapprox$40 K.  It is plotted on the T=30 K
curve at $z=1.9$.  The remaining sources other than SMM J123727+621042
are plotted with redshift upper limits well within the epoch of
reionization.  However, as with sources plotted on the T = 30 K curve,
the 2$\sigma$ limits cannot be accommodated for $z\gtrapprox 2.5$.
That 7 of 9 of the 450 $\mu$m sources in the GOODS-N field require ice
to explain the 850 $\mu$m non-detections without recourse to any other
constraint suggests strongly that the possible effects of ice emission
must be considered when interpreting submillimeter observations.  Ice
dramatically changes redshift limits: by a factor of 20 for SMM
123631+620657.

From the other direction, for the sources with 850 $\mu$m
detections and 450 $\mu$m upper limits that also have
estimated redshifts (\citealt{borys04}) we may also begin to
constrain the presence of ice.  14 upper limits are plotted
in Fig. 5 as open circles for sources where a redshift
estimate is given.  The open diamonds are also upper limits
plotted (but without arrows to avoid crowding) at the lower
limits to their redshift given by \cite{borys04}.  For
redshift estimates with $z<3$ the upper limits are
constraining on the likelihood that optically thin ice
contributes to the FIR opacity: at the 2$\sigma$ level
(arrow heads) little T$>$30 K optically thin ice seems to be
needed.  At the 3$\sigma$ level 4 of 11 sources could be
consistent with a contribution of optically thin ice
emission to the source SEDs.  One source with a redshift
lower limit ($z>2.7$; SMM J123652+621225) needs to be
optically thick and cool (T $\sim$40 K) or at a higher
redshift as already noted by \cite{borys04}.  The remaining
sources with with redshift estimates above 3 or redshift
lower limits are not constraining on the presence of ice.
The present considerations suggest that the 450 and ($z<3$)
850 $\mu$m detected sources differ physically in that the
former have optically thin ice emission that dominates their
FIR opacity while the latter may be optically thick at
850/(1+$z$) $\mu$m or that refractory dust dominates their
FIR opacity or both.  The average 450-to-850 $\mu$m ratio
for the 850 $\mu$m source calculated from the stacked 450
$\mu$m and the average of the 850 $\mu$m flux densities
reported by \cite{pope} is 0.5 which we have plotted as a
small open square at $z=4.25$.  The plotted redshift would
be consistent with T=30 optically thick material or T=20
optically thin ice or $\beta=2$ dust.  If it is true that
850 $\mu$m sources are on average fainter at 450 $\mu$m than
at 850 $\mu$m then it seems at though many of the 850 $\mu$m
sources must lie at $z>4$ and ice may be needed to make them
detectable at 850 $\mu$m with ULIRG luminosities (see end of
Sec. 4.2).  We do worry, however, that the stacked 450
$\mu$m flux is underestimated owing to the smaller 450
$\mu$m beam or that the data were normalized prior to
stacking making the reported units ambiguous.  QSO
J1148-5251 is also plotted (small filled circle).  As
discussed above, it seems to require ice emission.

Just as detected 850 $\mu$m sources must be at least
ultraluminous, so also must the detected 450 $\mu$m sources
be hyperluminous (L$>10^{13}$ L$_\odot$).  In Fig. 6 we plot
the models shown in Fig. 5 giving their 450 $\mu$m
brightness as a function of redshift under the following
assumptions about luminosity: The models for refractory
dust, and ice at T=30 and 50 K are ten times more luminous
than ULIRG IRAS 00188-0856 (Fig. 1), the model for T=150 K
ice is 10 times more luminous still, and we have also
plotted the complex amorphous ice model for QSO J1148+5251
at it's observed luminosity (short dashed line and filled
circle; $\sim 10^{13}$ L$_\odot$;
\citealt{beelen}).  The brightnesses of the 450 $\mu$m
sources given by \cite{pope} lie in the range 77-to-291 mJy
so that the plotted models would need to be scaled up by a
factor of a few to cover this range fully unless the sources
are at $z<0.2$.  For example, one of the brighter sources
(SMM J123631+620657; 263 mJy) would have a luminosity of
6$\times10^{14}$ L$_\odot$ for the T = 150 K ice model if
located at $z=9.5$, a factor of three higher than the most
luminous hyperluminous source listed by \cite{verma}.  It's
luminosity would be 10 times lower for the T = 30 K ice
model located at $z=2$, near the middle of the distribution
of luminosities in their list.  From Fig. 5 we estimate that
the 450 $\mu$m sources would populate the redshift ranges
1--3 and 5--10, with the latter range having a larger
comoving volume by a factor of 1.7.  If ice is only
important close to reionization so that the sources were all
at 9.2$<z<10.2$ (see solid line in Fig. 6) then the
luminosities require about a tenth of the stellar mass of
M87 per source to be involved in star formation.  The space
density would be 2.5$\times 10^{-5}$ Mpc$^{-3}$, comparable
to the local density of clusters of galaxies
(\citealt{bahcall}) and the luminosity density would be
$\sim5\times10^9$ L$_\odot$/Mpc$^3$ (comoving), about a
factor of 100 higher than the local FIR value
(\citealt{saunders}).  This would then imply a minimum of
10$^3$ recombinations per proton
if the 450 $\mu$m luminosity traces a factor of ten smaller
ionizing luminosity: about 10$^2$ more recombinations than for
the lifetime of a typical \ion{H}{2} region which seems too high
for reionization unless dust competes efficiently for
reionizing photons.

On the other hand, there are six 850 $\mu$m sources with flux
densities above 20 mJy listed by Pope et al. (2005) which
could be close to hyperluminous (see Fig. 7) so that we need
only postulate a similar population which is (thus far)
invisible a 850 $\mu$m owing to the effects of ice and thus
expect to find these sources at a range of redshifts.

There is very clear evidence of ice rather than dust providing the FIR
opacity in 7 of the 9 450 $\mu$m sources considered here.  But, the
redshifts may or may not be high and additional photometry is needed
to constrain the redshifts.  

\section{HDF 850.1 and ice}

\subsection{$z=5$}

The search for counterparts at other wavelengths to the brightest 850
$\mu$m source in the Hubble Deep Field has been arduous but
\cite{dunlop} make a convincing case for detections in the radio and
near infrared (NIR) and summarize submillimeter observations.
Modeling (\citealt{aretxaga}) of these observations leads them to
conclude that the redshift of HDF 850.1 lies in the range 4.1$\pm$0.5.
Since the SED of a high $q$ ULIRG (IRAS 08572+3915) is included in the
20 templates in the model input, the effects of the change in the
radio-FIR relation for ULIRGs compared to lower luminosity galaxies
noted in the introduction should be partly represented in their
estimate.  Similarly, the suggestion of ice emission in the SED of
IRAS 14348-1447 (Fig. 3) could also influence the estimate if the
photometric data were used directly rather than fitted with a modified
blackbody.  In general, reliance on modified blackbodies will give
lower estimated redshifts for cooler sources when more than one point
on the Rayleigh-Jeans curve is available, as is the case for HDF
850.1, but the addition of an ice emission feature can increase the
estimated redshift even for a cool source.

As one example, the bow-tie shown in Fig. 1 on the FIR continuum
corresponds to the range of redshift one would calculate for HDF 850.1
based on two estimates of its 850 $\mu$m flux density
(\citealt{dunlop}; \citealt{wang}) and the 1.3 mm flux density
reported by \cite{downes}.  That is 5.4 $<z<$5.6.  It should be noted,
even before describing the model upon which this is based, that a
detailed comparison between the SED shown in Fig. 1 and the optical
and NIR limits and detections for HDF 850.1 is not possible owing to
insufficient wavelength coverage for IRAS 00188-0856.  See
\cite{dunlop} fig. 6 for a comparison with data from two ULIRGs and a
luminous infrared galaxy calculated for a range of redshifts.
However, the effects of far infrared ice emission can be demonstrated
with this example.  The model, the {short-dashed} line in the
lower panel of Fig. 1, is based on curves given by \cite{aannestad}
which are the results of calculations of the optical properties of ice
mantled crystalline silicate grains.  And, it is crystalline water ice
in the mantles.  We chose the mantle-to-core volume ratio = 5.6 curve
to most closely match the ratio of 3 $\mu$m ice to silicate optical
depth used in the model shown by the solid lines in Fig. 1.  For a
number of reasons this choice is probably not fully constrained by this
ratio: the optical depth to which the silicate absorption is sensitive
may be larger than for ice; an odd shift in the wavelength of the
silicate feature may indicate a pyroxene composition with a different
feature-to-continuum optical depth ratio than usual; the crystalline
silicate in the \cite{aannestad} calculation is somewhat dissimilar to
astrophysical silicate models.  However the first two effects may be
compensating so that as a rough abundance analysis, the expectation
that ice would dominate the FIR opacity in this source seems well
founded.  This condition, and the need for the ice emission to be
moderately optically thin are really the only two requirements for ice
to strongly affect redshift estimation.  The optical depth of the
model of far infrared emission shown in the lower panel of Fig. 1 is
set to correspond to the 3 $\mu$m ice optical depth used in models
shown by the solid line in the upper panel (where ice and refractory
material contribute equally to the absorption optical depth at 3.1
$\mu$m).  The temperature of the far infrared emission model is set to
roughly reproduce the IRAS 60 and 100 $\mu$m measurements.  This temperature also accentuates
the effects of the broad 150 $\mu$m feature over which the HDF 850.1
data are displayed.  The range of redshifts needed to place HDF 850.1
on this model falls outside of the range estimated by \cite{dunlop}
and gives a first indication that the effects of ice on photometric
redshift estimation (as opposed to just limits) can be significant.

\subsection{$z=13$}

As noted in the introduction, using local ULIRGs to
understand submillimeter sources is a sensible approach if
their properties are not strongly affected by cosmic
evolution.  However, at higher redshift, this proviso seems
fairly likely to break down.  Many local ULIRGs appear to be
produced in the mergers of gas rich spiral galaxies and it
has been argued (\citealt{mihos}) that the dynamical
stability provided by bulges may preserve gas reservoirs for
the final merger stages associated with some ULIRGs.  Owing
to the finite lifetime of the universe, there are earlier
periods when the precursor systems may not resemble present
day precursors in their dynamical state or star-to-gas mass
ratio.  It is not at all clear that seed SMBHs
(\citealt{osterbrock}) would be present at sufficient mass
to allow accretion to be an energetically important or
dominant power source as suspected for local ULIRGs ({\it
e.g.} \citealt{idm}).  Thus, in what follows we assume star
formation as the main power source at higher redshift.  A
sketch of how crystalline ice might exist in a primordial
galaxy environment is also given to explore the plausibility
of a very large effect of ice on redshift estimation.  It is
to be noted that giving up a central power source such as
would be provided by an AGN would usually mean that ice
mantled grains would contribute little to the FIR emission
since most of the energy that powers FIR emission would be
absorbed in reactive photodissociation regions (PDRs) rather
than in molecular clouds where ice is thought to exist
unless an improbable scenario involving an overabundance of
protostars such as that proposed by \cite{roussel} but
previously discounted by \cite{dww} is invoked.  Below we
argue for such an overabundance.

In Fig. 7 we show the observations of HDF 850.1 at $z=12.6$ This is 18
confidence intervals beyond the \cite{dunlop} estimate.  So far as we
can tell, neither the new $Z$ band upper limit reported here, nor the
released {\it Spitzer} IRAC and MIPS GOODS images (\citealt{mark})
rule out either the model shown in Fig. 7 or the analysis of
\cite{dunlop} so long as a reddish ULIRG (see \cite{dudley} for a
discussion of 12-to-60 $\mu$m flux density ratios in ULIRGs) is used
as a basis for the \cite{dunlop} analysis.  \cite{dunlop} argue that
the lensing source (solid diamonds in Fig. 7) is evolved, in which
case much of the tentative observed frame 24 $\mu$m flux density could
be owing to HDF 850.1 but little of the 8 $\mu$m flux density.  At
$z=4.1$, Mkn 231 and 273 could be too blue to account for this level
of emission whereas Arp 220 or IRAS 17208-0014 would not violate this
datum if treated as an upper limit.  On the other hand, at $z=1.1$ the
chances of finding a younger elliptical galaxy to act as a lens are
increased so that some or all of the observed 24 $\mu$m flux density
could be owing to the lensing galaxy.  This would still not rule out
the \cite{dunlop} redshift estimate for HDF 850.1 but would
accommodate the present estimate under the assumption of the plotted
starburst model.  The shaded region in Fig. 7 shows a possible range
for the spectral shape of the lens source emission.  The source
labeled 1b in fig. 8 of \cite{wang} may also contribute to the 24
$\mu$m emission though only slightly to the IRAC data. If it is
similar to NGC 253, and at a redshift of $z_{\rm ph}=1.76$
(\citealt{fs}) then up to $\sim$30\% of the 24 $\mu$m flux density
might be attributed to this source but again negligibly at 850 $\mu$m.

The model for HDF 850.1 shown in Fig. 7 ({thick gray}
line) is similar to that shown in Fig. 1 in that it has the
same mantle-to-core volume ratio and optical depth however the
temperature is higher as it must be owing to the higher CMB
temperature at $z=12.6$.  Whereas in Fig. 1, the 150 $\mu$m
ice feature spectral structure influences the redshift
placement of the HDF 850.1 data, in Fig. 7, shorter
wavelength spectral structure also owing to ice is
influential.  Interestingly, assuming that there is a
foreground neutral intergalactic medium as a result of
incomplete reionization (\citealt{spergel06}), one would expect
Gunn-Peterson saturation or a Ly$\alpha$ blackout to occur
in the $H$ band for $z=12.6$ the shortest wavelength band for
which HDF 850.1 is detected.  This is represented by the
change from black to gray in the thin line using a Heaviside
function as multiplier.  Here the {thin line} is a young
starburst spectrum generated using Starburst99
(\citealt{claus}).  It is notable that taking the onset in
redshift of the Gunn-Peterson trough to be $\sim$6
(\citealt{fan}) absorption at the Lyman limit at $z<12.6$
easily overlaps Ly$\alpha$ absorption between $z=6$ and
$z=9.2$.  One expects pretty thorough opacity shortward of
Ly$\alpha$.

While the choices of parameters of the starburst model are meant only
to be suggestive, for the model shown in Fig. 7, the free-free
emission expected based on the rate of production of ionizing photons
would explain about 20\% of the observed radio emission (following
\citealt{nick}) while retaining the starburst model as it is and
correcting for sufficient extinction to account for the model infrared
emission would over produce the radio emission so that a lower
contribution of stars which produce ionizing photons to the observed
frame $K$' band flux might provide a more likely scenario.
Alternatively, invoking a chemical trigger for star formation
discussed below, it is conceivable that an object could be caught in a
phase when a sufficient number of stars are in an ultracompact
\ion{H}{2} region configuration that, given a sufficiently high
ionization parameter, higher than usually required if the gas-to-dust
mass ratio is high owing to low enrichment, free-free emission would
be suppressed ({\it e.g.} \citealt{bottorff}).

As noted in the introduction, sources at high redshift
should eventually drop out of 850 $\mu$m surveys.  In Fig. 8
we plot, for the same far infrared luminosity, the expected
850 $\mu$m brightness as a function of redshift for an Arp
220-like SED and the two distributions shown in Figs. 1 and
7.  The 30 K model is truncated at $z=11$ since the cosmic
background radiation temperature exceeds the model
temperature above this redshift.  As can be seen, the models
with ice select for high redshift sources assuming such
sources exist.  For the 30 K model the brightest (non-local)
source would be found near a redshift of 5 and for the 50 K
model near a redshift of 13.  Should ice play the role proposed here
in a moderate fraction of the detected 850 $\mu$m sources,
it would be of little surprise that the brightest source in
the HDF (850.1) would be among them.

\subsection{Why star formation?}

Our choice of presenting a model that uses star formation as
a power source may seem strange given that PDRs would be
inefficient at producing ice emission which we require to
allow, for example, HDF 850.1 to be detected at $z=13$.  One
phase of star formation that does allow MIR radiation to heat ice
is the protostar phase.  This phase is brief
compared to O star life times and thus would not typically
be expected to provide the dominant source of FIR radiation
in a galaxy.  Protogalaxies, however, are by definition
required to be young, so that while the argument concerning
the relative brevity of the protostar phase cannot be
ignored, it may be overcome if the star formation rate is
increasing rapidly with time.  And, it is just such sources
that one might expect to be observed as they populate the
top of the infrared luminosity function since they would be
the most efficient at converting radiation to FIR
wavelengths.

A rapidly increasing star formation rate might be expected
in a situation where the gas cooling efficiency is increasing
owing to ongoing initial enrichment.  A positive feedback
can lead to an exponentially increasing system response.  
The number at the right hand of Fig. 8 shows the 850 brightness (in
mJy) of the three SED shapes considered here if we adopt the lensing
amplification proposed by \cite{dunlop} for HDF 850.1 for a source
with the FIR luminosity of IRAS 00188-0856.  The increase in
luminosity required to match the \cite{hughes} estimate for HDF 850.1
would be a factor of 2.3 for the dashed curve taken at a redshift of
13.  This would then correspond to $\sim 4\times 10^{12}$ L$_\odot$
({\it e.g.} Fig. 7) or 10$^7$ O stars and 1.5$\times 10^{11}$ G0 stars
for a Salpeter initial mass function (m$^{-2.35}$), or roughly a
galaxy's worth of star formation.  This then would be the limit for
the continued exponential increase in the star formation rate: the
amount of available enrichable gas.  Night et al. (2005) find 4
objects with $10^{11}$ M$_\odot$ in stars at $z=6$ in their largest
cosmological simulation (142 Mpc; large enough to contain $\sim$1
ULIRG at $z=0$).  Thus, there is some suggestion that relatively
massive objects (or perhaps regions that will become objects) are
forming their stars a few dynamical timescales earlier than this.
Notwithstanding the existence of QSO J1148+5251 and other high
redshift quasars, it seems to us that there is a need to account for
massive sources at intermediate redshift which last formed stars near
$z=13$ (\citealt{alan}; \citealt{jimenez}) as well as the apparent
monotonic increase in the comoving rate of gamma-ray bursts out to
$z=7$ (\citealt{chuck}).

However, in the low-to-moderate redshift universe, we would
usually take ice emission as an indication of a buried compact
power source at least at ULIRG luminosities, and, should blackhole 
growth commonly reach $\sim 10^7$ M$_\odot$ by $z=13$, as must happen
at least occasionally given the existence of  QSO J1148+5251, then
our low-to-moderate redshift explanation could also apply during
reionization.

\section{Summary and prospects}

Far infrared ice emission may be crucially important to understanding 
the far infrared emission of sources at high redshift and may strongly affect 
the range of redshifts estimated for submillimeter sources, even placing
them at reionization redshifts.  Observations of infrared ice emission 
are definitive in Galactic protostars and the winds of evolved stars
but are scant and not fully persuasive for the local analogs of the 
submillimeter sources, the ULIRGs. This, despite mounting evidence that
ice is involved in absorbing much of the energy that is finally emitted 
in the far infrared in a substantial fraction of ULIRGs. 

We find stronger indications that ice emission is important at high redshift.
The data for QSO J1148+5251 as well as the 450 $\mu$m sources cataloged
by \cite{pope} are both strongly suggestive. We feel that 
further investigation along these lines could be quite fruitful.

We've considered two models where the presence of far-infrared
emission features owing to ice in the SED of HDF 850.1 would place it
at redshifts larger than the range estimated without ice by
\cite{dunlop}.  In the first, the use of crystalline ice at 150 $\mu$m
as a substitute for amorphous ice is probably justified, subject to
further laboratory investigations of amorphous ice.  In the second,
consideration of a first chemical enrichment mechanism suggests that
crystalline ice could be present at early times while energetic and
timescale considerations may be compatible with protostars in
protogalaxies producing the sort of features we consider here.  In the
second model, a Lyman $\alpha$ blackout would be consistent with the
NIR photometry for HDF 850.1.  However, we do not prefer this model
and its higher redshift over the redshift range estimated by
\cite{dunlop} or $z\sim 5.5$ from our first model.  What we find to be
important is that far infrared ice emission could have such a dramatic
effect on the estimated redshift.  It should be noted that redshifts
estimated from the radio-FIR relation could also turn out to be lower
when ice emission is invoked since a large submillimeter-to-radio flux
density ratio is easier to produce with high $\beta$ at some
redshifts.  So, the main conclusion of this work is that ice broadens
the range of redshifts to be considered for a given submillimeter
source and that ice is quite likely to be important at high redshift.

Some astronomical observations
which could aid progress on open issues are:
\begin{itemize}
\item Continuum observations at 3 mm of QSO J1148+5251 may confirm 
ice emission in this source. 
\item The Atacama Large Millimeter Array  may be able to isolate
the transition from dust to free-free/synchrotron emission for sources
of this type and allow more reliable redshift estimates.
\item Investigating the role of ice emission locally could be taken up using
SOFIA or Spitzer. 
\item Very low resolution H-band spectroscopy might test the
possibility of a Lyman $\alpha$ blackout for HDF 850.1.
\item Molecular studies which target H$_2$O and OH maser emission may
give definite redshifts using a new generation of radio telescope 
arrays even if the continuum is below detection limits.
\end{itemize}

Colleagues have commented helpfully at various stages of this
work. Some are: S. Chubb, E. Dwek, J. Fischer, P. Gerakines, A. Li,
M. Moore, L.~J. Rickard, E. Shettle, and H. Smith and an anonymous
referee.  Research in infrared astronomy and particularly the
astrophysics of celestial backgrounds at the Naval Research Laboratory
is supported by the Office of Naval Research (USA). This publication
makes use of data products from the Two Micron All Sky Survey, which
is a joint project of the University of Massachusetts and the Infrared
Processing and Analysis Center/California Institute of Technology,
funded by the National Aeronautics and Space Administration and the
National Science Foundation.  Use is also made of the USNOFS Image and
Catalog Archive operated by the United States Naval Observatory, the
archive of results from the ESA sponsored {\it Infrared Space
Observatory} mission, the NASA/IPAC Extragalactic Database operated by
JPL and Caltech, and data from the GOODS project from NASA's {\it
Spitzer} and {\it Hubble Space Telescopes}.

\clearpage
\begin{figure}
\parbox{6.5in}{\caption{ Upper Panel: Data from
\protect\cite{masa} is reproduced for IRAS 00188-0856
together with a model (solid line) subsequently employed in
the lower panel.  The model consists of an $\alpha = 2$
(flux density $f_\nu \propto \nu^{-\alpha}$) power-law
continuum to which absorption is applied together with an
unabsorbed component which is a combination of an archival
spectrum of \hbox{M 82} (ISO SWAA11600319) and a
blackbody of temperature 2500 K to represent additional
stellar photospheric emission (displayed in the lower panel
long-dashed line).  The two absorption components used are:
(1) a volatile admixture (H$_2$O:CO=2:1) of amorphous ice
measured at 10 K (\protect\citealt{hagen}) and (2) the
refractory EURECA-B curve presented by
\protect\cite{greenberg}.  The effects of absorption by the
refractory material are shown with the dot-dashed line and
those of the combined volatile and refractory absorption
with the dashed line.  The volatile and refractory
components have approximately equal contributions to the
absorption at 3 $\mu$m but the refractory component
contributes little at the short wavelength edge suggesting
that a volatile component is required.  Lower Panel: The
same model (solid line), now adjusted to lower optical depth
in the refractory component to better match the 6 $\mu$m
absorption and with a smaller additive M 82 component is
applied to longer wavelength data.  Here, the effects of
silicate absorption have also been included based on fig. 5
of \protect\cite{dww} but oddly shifted to the red by about
0.6 $\mu$m at 10 $\mu$m, perhaps owing to differing composition.  The inset shows the effects of
adding the spectrum of \hbox{M 82} as in the upper panel and
without this component in the model at all (bounding dotted
lines).  The spectral resolution of the model matches that
of the data in the inset.  The short-dashed line is an
emission model described more fully in the text (Sec. 4.1).  Data
sources for IRAS 00188-0856 in descending order in the
legend are from \protect\cite{masa}; \protect\cite{monet};
\protect\cite{cutri}; \protect\cite{iras}; and
\protect\cite{charmandaris}.  R and K' data are in rough
agreement with the photometry given by \protect\cite{kim2}.
}}
\end{figure}

\begin{figure}
\caption{ Highly processed data products from the ISO
Archive (hpdp\_05100540\_2.tar and  hpdp\_65000608\_2.tar with
minimal ISAP post processing) are compared with a 43 $\mu$m
ice absorption feature derived from the spectrum of HD
161796 (also from the ISO archive: SWAA52100537). The diamonds and circles are data for Mkn 231 and IRAS 17208$-$0014 respectively.  The solid lines are derived from the ISO SWS spectrum of HD 161796 using a single temperature (140 K) to estimate emissivity after the manner of \protect\cite{ra84} and then converted to  absorption. The inset is described in the text.}
\end{figure}

\begin{figure}
\caption{
Observational data for IRAS 14348-1447 (filled
circles) are plotted at their rest wavelength in Jy as reported by
\protect\cite{klaas}.  Horizontal bars indicate filter passbands.  The 
solid line shows an optically thin (in the FIR) ice emission model
whose $\beta$ parameter (100 and 300--900$\mu$m) is $\sim 2.3$ as
compared to $\sim 1.9$ for a bare (refractory) core.  Adjustment of
the core to smaller $\beta$ would result in a smaller $\beta$ for the
mantled grain as well.  In fitting the model data at 95 and 450 $\mu$m
were matched and various mantle-to-core volume ratios were compared
with photometry between these wavelengths.  The dashed line
reproduces the fit proposed by \protect\cite{klaas} which is quite optically
thick in the FIR and has $\beta$=2.
}
\end{figure}

\begin{figure}
\caption{
Observational data for QSO J1154+5251 (filled
circles and arrow) are plotted at their rest wavelength in mJy as reported by
\protect\cite{berto1} (3 mm); \protect\cite{berto2} (1.2 mm); \protect\cite{robson} (850 and 450 $\mu$m); \protect\cite{beelen} (350 $\mu$m).  The lines represent emission models.  The dashed and dot-dashed are single temperature models using refractory and ice mantled dust respectively while the solid line two temperature ice model with colder ice obscuring warmer ice.
}
\end{figure}

\begin{figure}
\caption{The observed frame 450-to-850 flux density ratio calculated
for a series of emission models is plotted as a function of redshift
together with 9 observed lower limits (large filled symbols), 34
observed upper limits (medium open symbols), and two detections
(smallest symbols). Individual 450 $\mu$m source and detection
identifications are given in the legend.  The models, also in the
legend, are more fully described in the text.  }
\end{figure}

\begin{figure}
\caption{Brightness in mJy at the observed wavelength 450 $\mu$m is
plotted as a function of redshift for a set of models.  Four of the
models are carried over from Fig. 5: The solid line represents 150 K
ice, the triple dot dashed line, 50 K ice, the dot dashed line 30 K
ice and the long dashed line 60 K refractory dust.  A fifth model
(short dashed line) is taken from the solid line in Fig. 4: 71 K ice
is obscured by 24 K ice.  Infrared luminosities for the models are
constant as a function of redshift and are $2\times 10^{13}$
L$_\odot$: long-dashed, dot-dashed, and triple-dot-dashed lines,
$2\times 10^{14}$ L$_\odot$: solid line, and $1\times 10^{13}$
L$_\odot$: short-dashed line.  We remark that just as standard dust
shows roughly redshift independent brightness at 850 $\mu$m, so too
does T=50 K ice (triple dot dashed line) at 450 $\mu$m.  The filled
circle is QSO J1148+5251. }
\end{figure}

\begin{figure}
\caption{ Observational data for HDF 850.1 (filled circles) are
plotted for a rest wavelength and luminosity scale corresponding to a
redshift of 12.6 and accounting for the effects of gravitational
lensing by reducing the apparent luminosity by a factor of 3.  The
data, except for the $Z$ band limit, are reported or summarized in
\protect\cite{dunlop}.  The $Z$ band limit is measured from GOODS ACS
data (\citealt{giavalisco}) by subtracting the image of the lensing
galaxy from itself after a rotation of 180\degr.  The limit is
2$\sigma$ drawn from the weight image for an 0\farcs5 aperture
centered at the IRAM position (\citealt{downes}) and corresponds to
$Z-K'>3.5$ (Vega).  Data for the lensing source (filled diamonds)
should be read relative to the HDF 850.1 data (filled circles) as the
scalings in wavelength and luminosity do not apply.  Here the
$I-K^\prime$ photometry are from \protect\cite{dunlop} while the
3.6--24 $\mu$m points are measured from GOODS IRAC and MIPS images
(\protect\citealt{mark}).  The IRAC data are 6\farcs0 diameter
aperture photometry subtracting an equal area outer annulus.  The 24
$\mu$m datum is obtained through radial profile fitting after
subtraction of a confusing source and is perhaps a 2.5$\sigma$
measurement of 20 $\mu$Jy.  The dashed line is scaled ($z=1.1$ wrt
observed frame; \citealt{fs}) data for the elliptical galaxy NGC 1399
(\citealt{Persson};\citealt{athey}) and the triple-dot-dashed line is
for the galaxy NGC 3928 (\citealt{huchra}; 2MASS; \citealt{xilouris})
with the shaded region between them showing the age dependent range of
mid-infrared flux suggested by models for elliptical galaxies reported
by \protect\cite{temi}.  The models shown as solid lines are
described more fully in the text.}
\end{figure}

\begin{figure}
\caption{ The 850 $\mu$m brightness as a function of redshift is shown
for three calculated SEDs.  The solid line corresponds to the FIR SED
shown in Fig. 1. The long-dashed line corresponds to an optically thin
$T=45$, $\beta=1.2$ modified blackbody similar to the
\protect\cite{nick} fit for Arp 220.  The short-dashed line
corresponds to the FIR SED shown in Fig. 7 all normalized to the FIR
luminosity of IRAS 00188-0856.  The arrows for the solid, dot-dashed,
and dashed lines indicate redshifts for HDF 850.1 estimated using
Fig. 1, \protect\cite{dunlop}, and Fig. 7 respectively.  The numbers at
the right are a conversion of the vertical scale for the gravitational
lens amplification factor of three estimated for HDF 850.1.  }
\end{figure}

\clearpage
\begin{figure}
\vspace*{25cm}
\vskip-25cm\includegraphics[scale=.90]{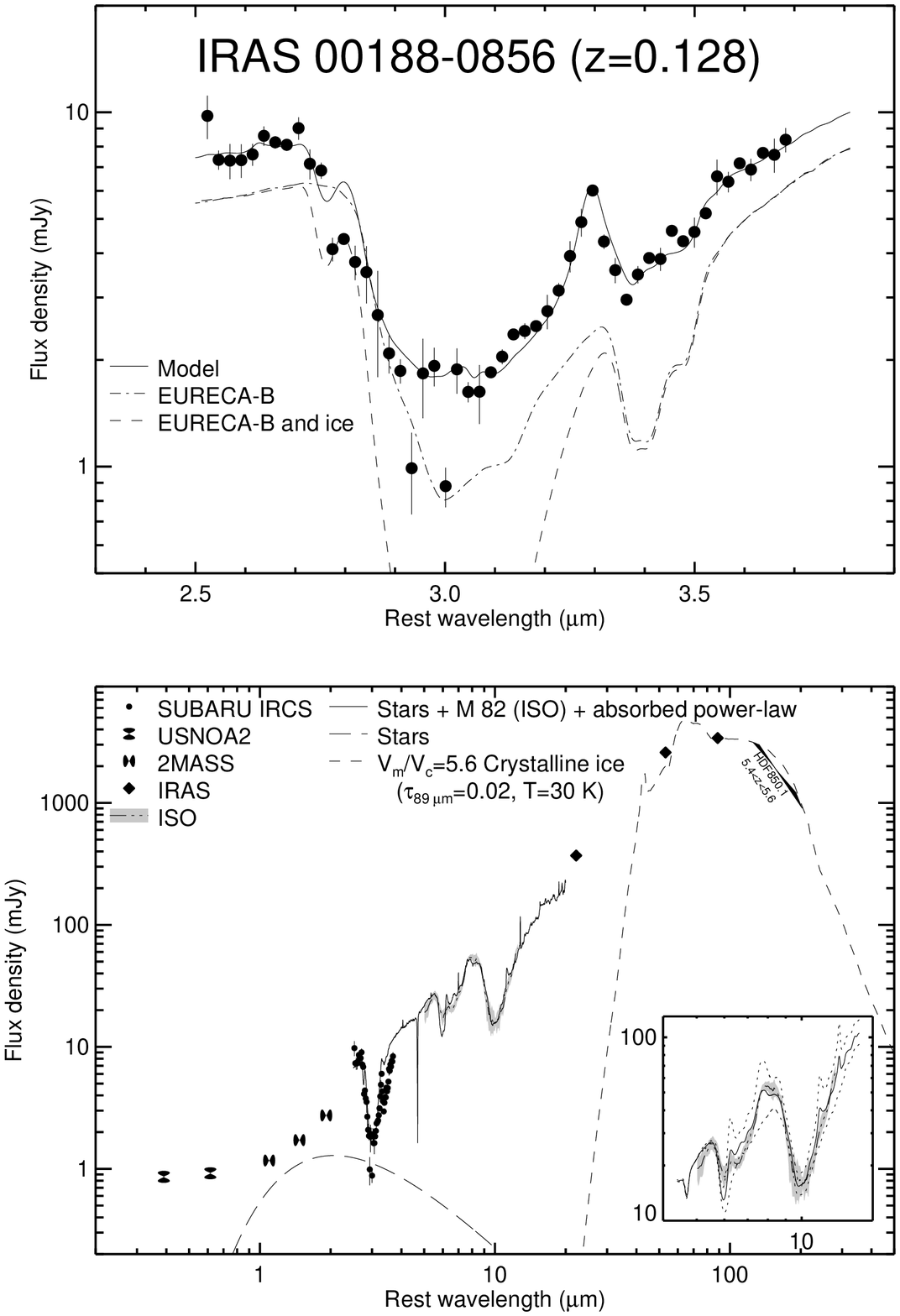}
\figurenum{1}
\caption{}
\end{figure}
\clearpage

\begin{figure}
\vspace*{12.5cm}
\vskip-12.5cm\includegraphics[scale=.70]{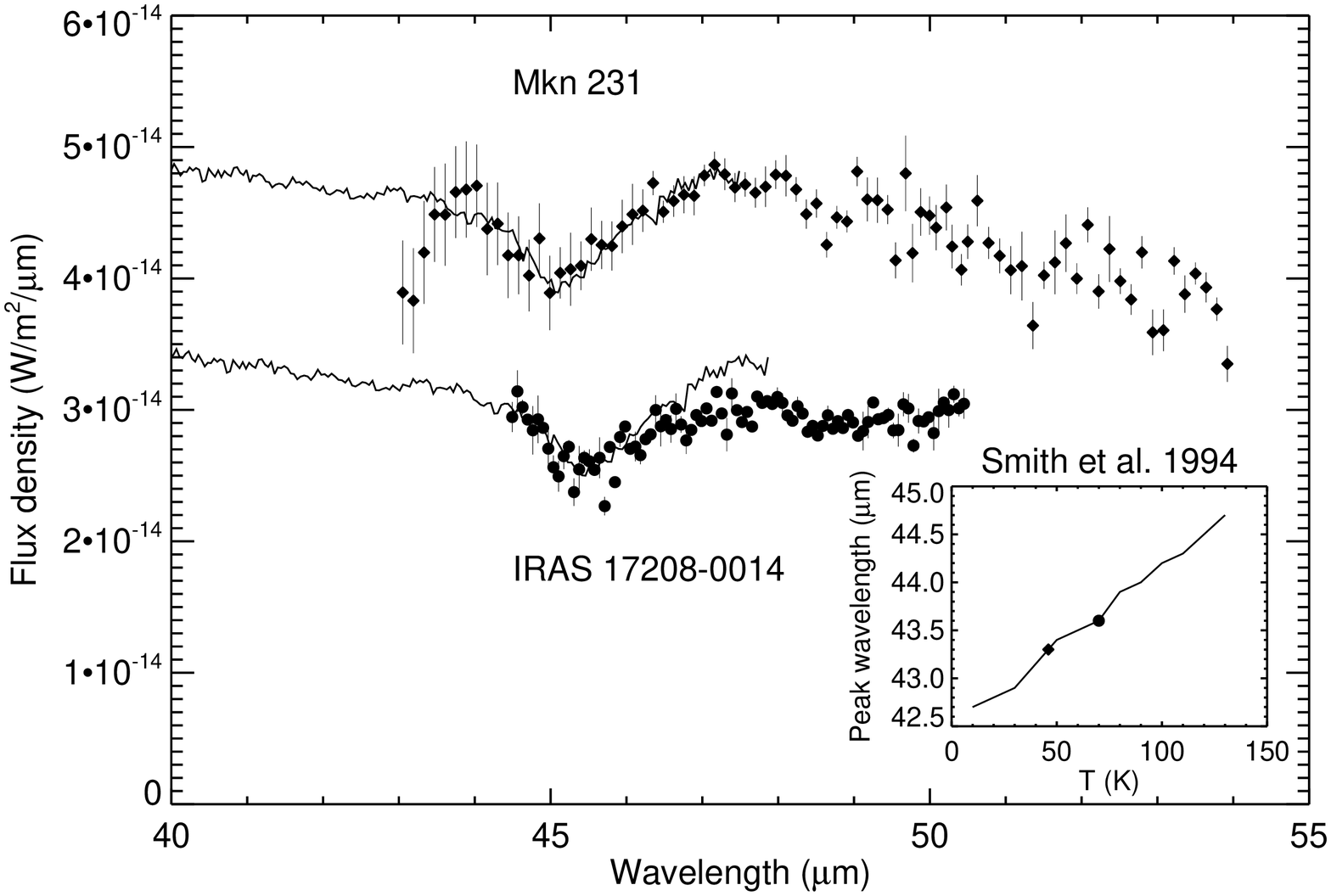}
\figurenum{2}
\caption{}
\end{figure}
\clearpage

\begin{figure}
\vspace*{12.5cm}
\vskip-12.5cm\includegraphics[scale=.70]{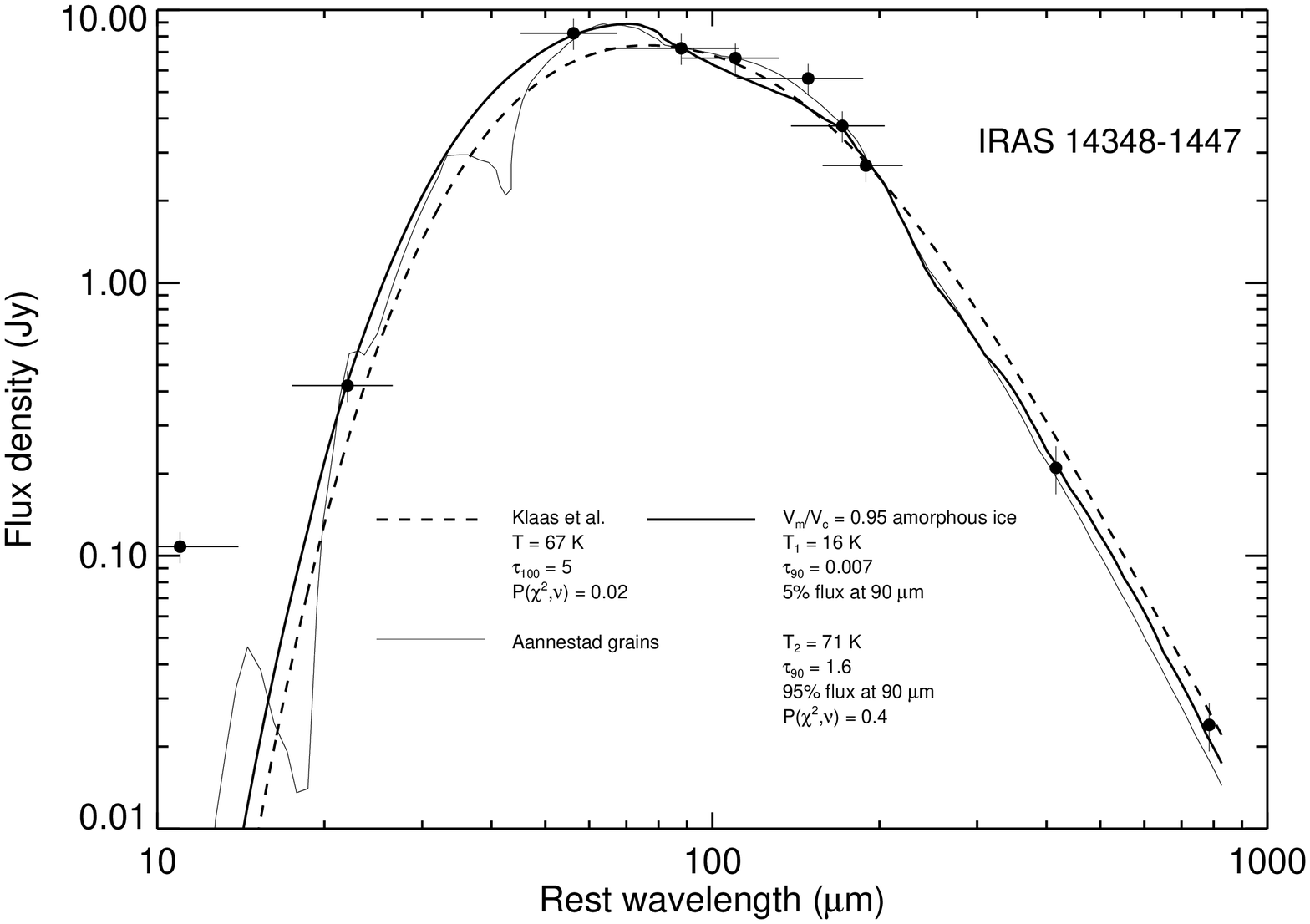}
\figurenum{3}
\caption{}
\end{figure}

\clearpage

\begin{figure}
\vspace*{12.5cm}
\vskip-12.5cm\includegraphics[scale=.70]{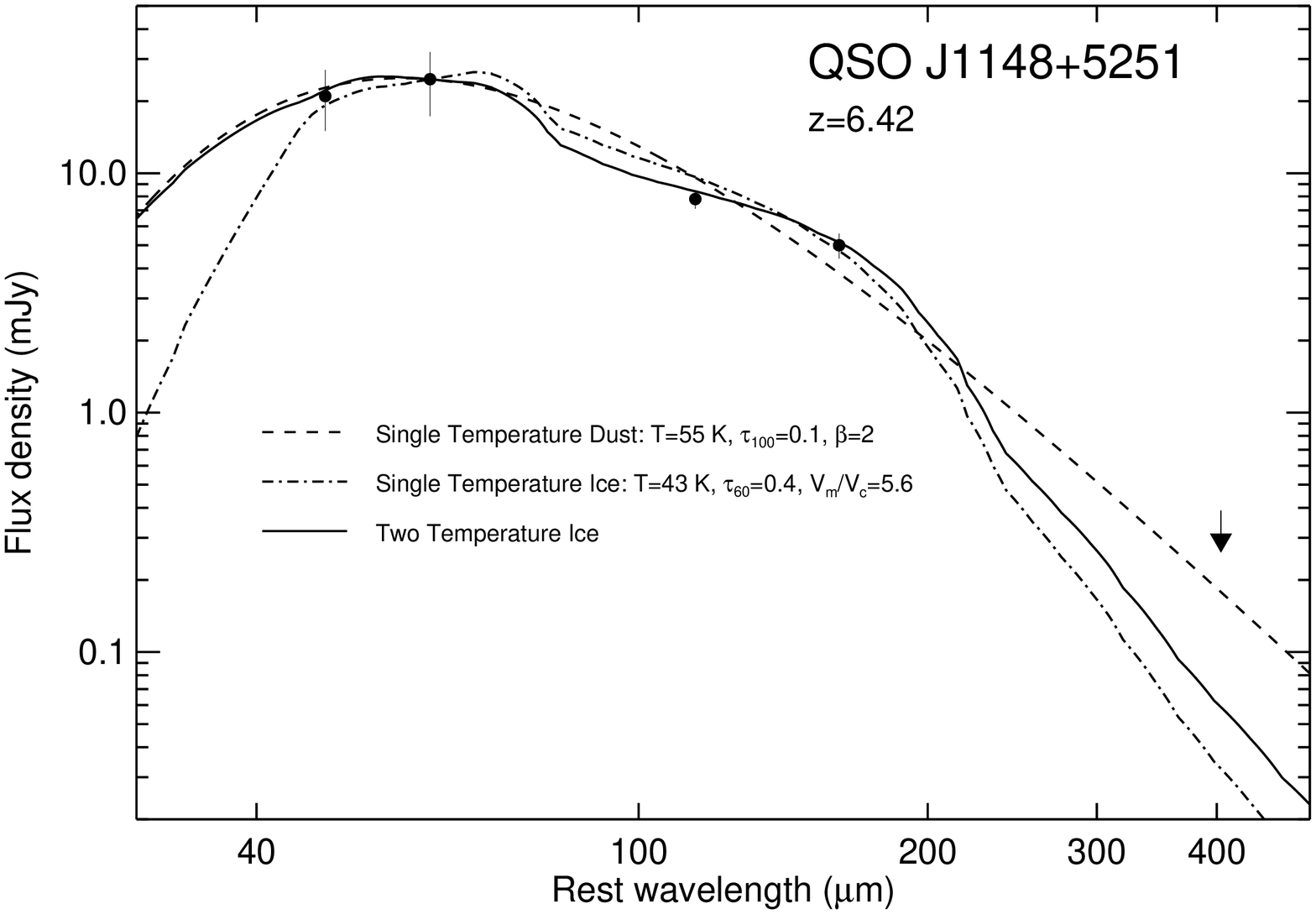}
\figurenum{4}
\caption{}
\end{figure}

\clearpage

\begin{figure}
\vspace*{12.5cm}
\vskip-12.5cm\includegraphics[scale=.70]{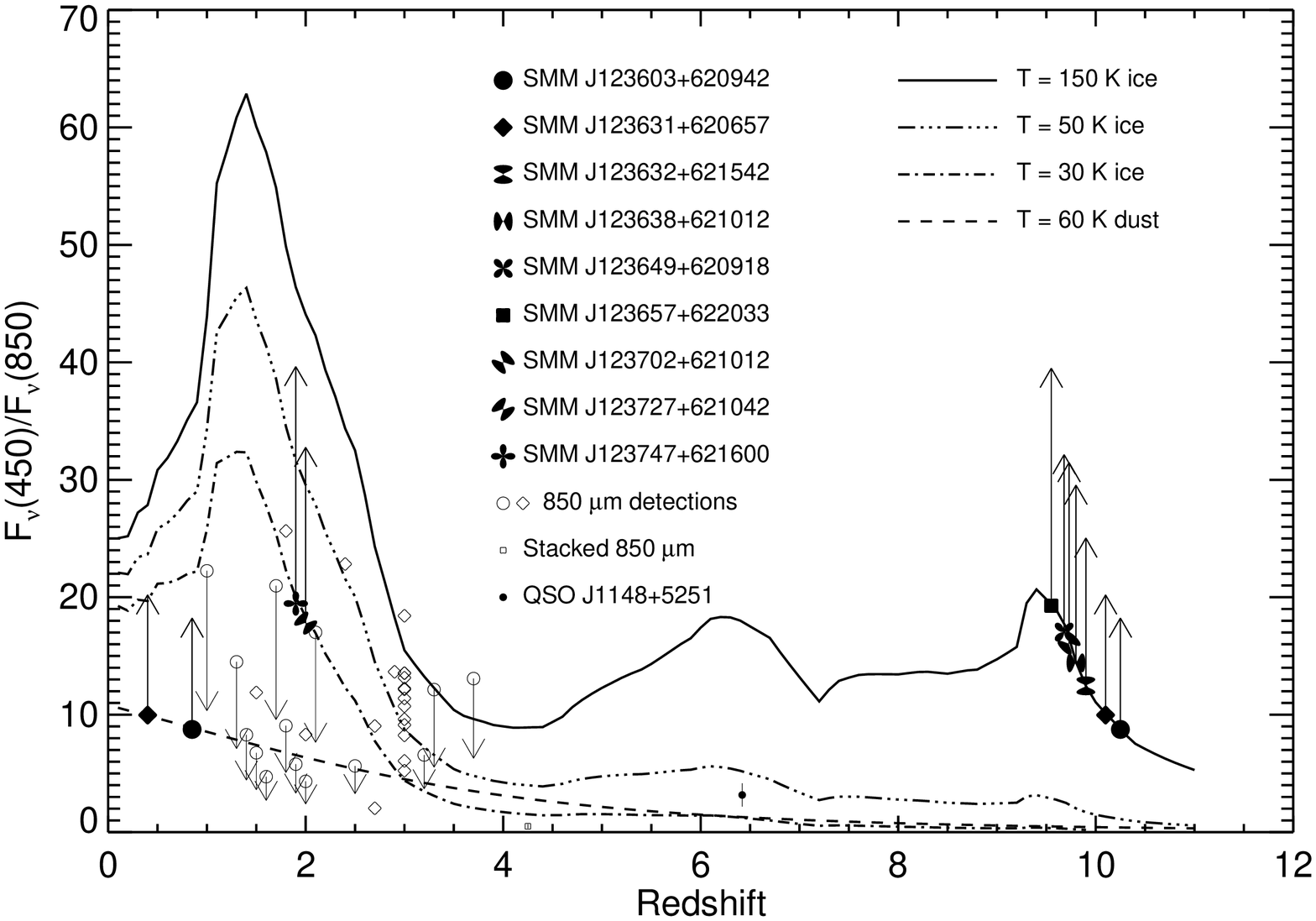}
\figurenum{5}
\caption{}
\end{figure}
\clearpage

\begin{figure}
\vspace*{12.5cm}
\vskip-12.5cm\includegraphics[scale=.70]{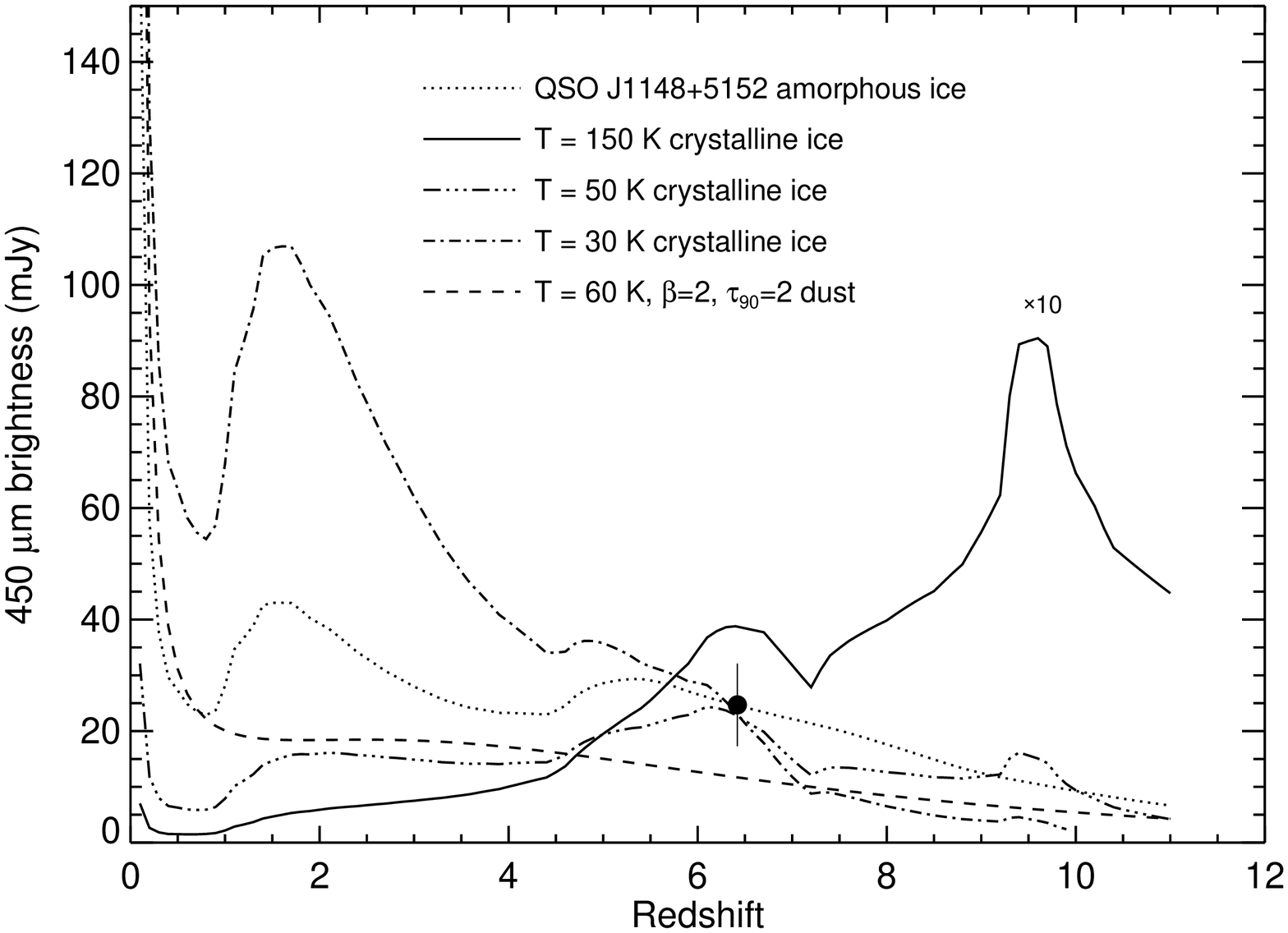}
\figurenum{6}
\caption{}
\end{figure}
\clearpage

\begin{figure}
\vspace*{12.5cm}
\vskip-12.5cm\includegraphics[scale=.70]{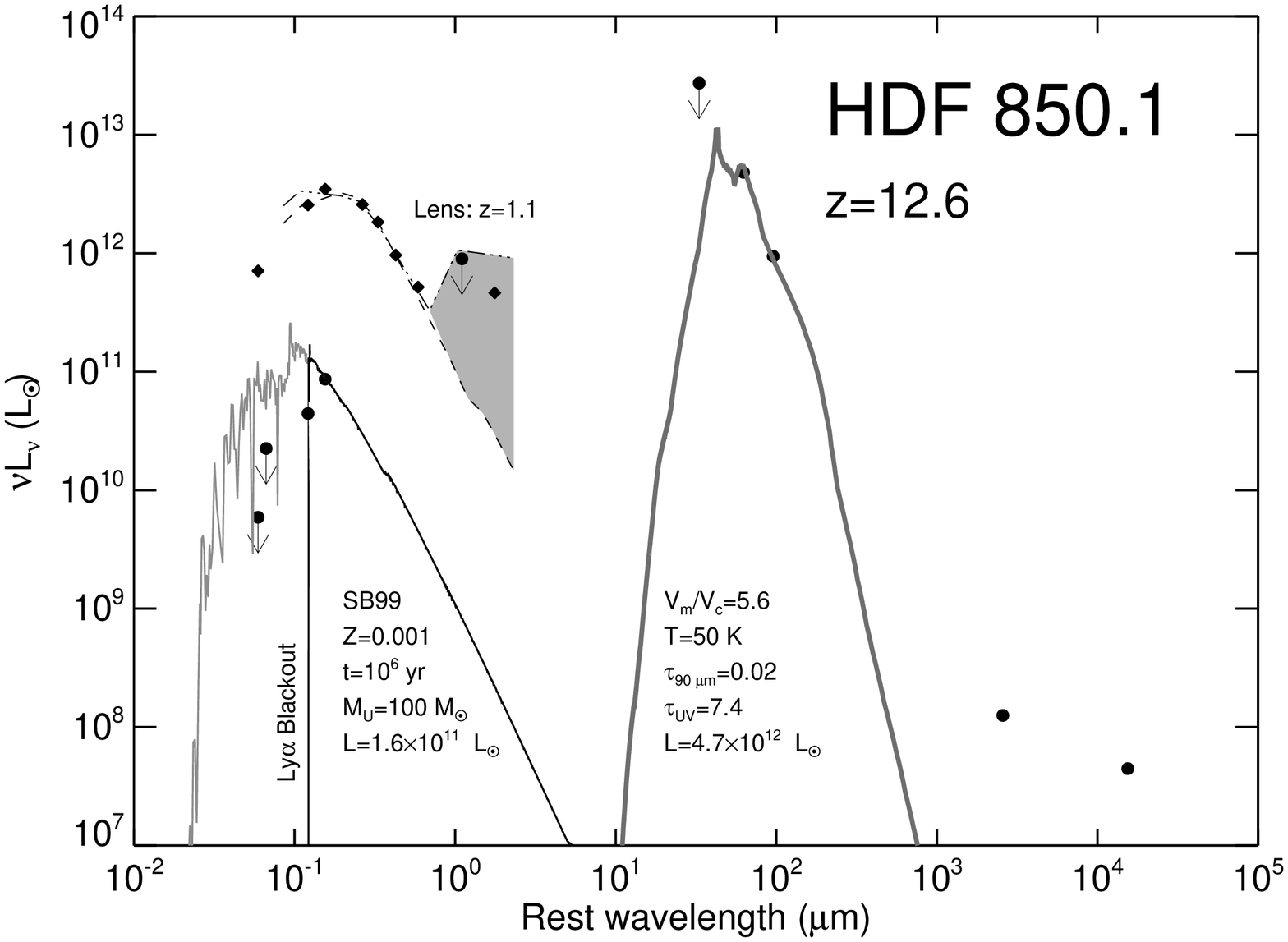}
\figurenum{7}
\caption{}
\end{figure}

\clearpage

\begin{figure}
\vspace*{12.5cm}
\vskip-12.5cm\includegraphics[scale=.70]{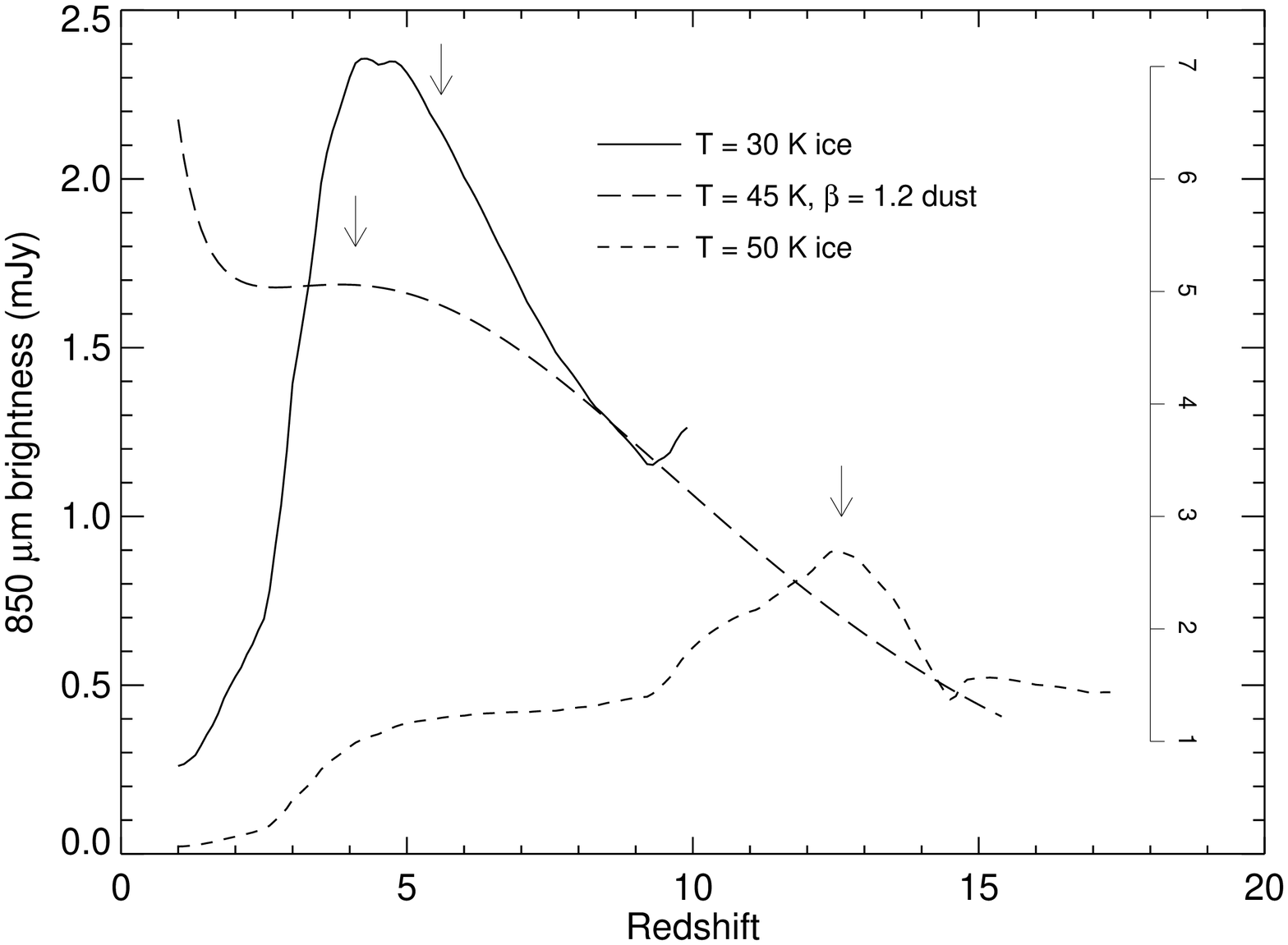}
\figurenum{8}
\caption{}
\end{figure}
\clearpage

\end{document}